# Magnetic properties of ruthenium dioxide (RuO$_2$) and charge-magnetic interference in Bragg diffraction of circularly polarized x-rays

S. W. Lovesey[1, 2], D. D. Khalyavin[1] and G. van der Laan[2]

[1]ISIS Facility, STFC, Didcot, Oxfordshire OX11 0QX, UK

[2]Diamond Light Source, Harwell Science and Innovation Campus, Didcot, Oxfordshire OX11 0DE, UK

**Abstract** Rutile-type RuO$_2$ likely supports a simple antiferromagnetic structure which can be verified by x-ray Bragg diffraction. Three magnetic motifs that do not break translation symmetry are explored in calculations of amplitudes suitable for diffraction enhanced by tuning the primary x-ray energy to a ruthenium atomic resonance. Coupling to x-ray helicity through a charge-magnetic interference is common to all motifs, together with magnetic and charge intensities in quadrature in the rotated channel of polarization. Necessary conditions for these diffraction phenomena are a centrosymmetric crystal structure, null magnetic propagation vector, and absence of a linear magnetoelectric effect. Published x-ray diffraction data for RuO$_2$ was analysed by the authors against a magnetic motif that does not satisfy the conditions. A polarized neutron study of antiferromagnetic domains can be achieved with a sample that meets the stated crystal and magnetic symmetries.

## I. INTRODUCTION

Crystals of ruthenium dioxide adopt a rutile structure with Ru ions in centrosymmetric octahedra, sharing edges along the c direction and corners in the a-b plane. Hitherto, the 4d transition metal oxide has not been popular in studies of materials with strongly correlated electrons and significant spin-orbit coupling. Not so with regard to technological applications and thin film devices, however [1-6]. Notably, RuO$_2$ is one of a few naturally highly conductive stoichiometric oxides and corrosion resistant. While ruthenium metal oxidizes very slowly, once formed RuO$_2$ is one of the most chemically stable oxides. The hardness of RuO$_2$ is just a little less than the hardness of fused silica. Antiferromagnetic long-range magnetic order is likely up to room temperature [7, 8]. Moreover, thin films of RuO$_2$ exhibit excellent diffusion barrier properties.

We forecast interference of charge and magnetic contributions to x-ray Bragg diffraction patterns for RuO$_2$ that is a signature of specific symmetry conditions [9, 10]. It is exposed by helicity in primary x-rays tuned in energy to an atomic resonance, and the chiral signature formed from a partial diffracted intensity can be different from zero. Notably, the signature is allowed with a centrosymmetric crystal, magnetic order that does not break translation symmetry (a propagation vector **k** = (0, 0, 0)), and anti-inversion ($\bar{1}'$) absent in the magnetic crystal class. The latter material requirement means charge-magnetic interference is forbidden in resonance enhanced x-ray diffraction by chromium sesquioxide or gadolinium tetraboride (see Appendix) that present a linear magnetoelectric effect. In addition to a chiral signature in diffraction different from zero, charge and magnetic intensities in the rotated channel of polarization are in quadrature. A calculation of the x-ray Bragg diffraction pattern of RuO$_2$ reported by Zhu *et al*. [8] yields a null chiral signature, because the calculation does not include the 90º phase shift between charge and magnetic contributions to x-ray scattering demanded by magnetic crystal symmetry. The symmetry error undermines the authors' claim of experimental evidence for collinear antiferromagnetism in RuO$_2$ [8]. Allied to charge-magnetic interference in diffraction of helical x-rays is a spontaneous Hall effect [11, 12] and a magneto-optical effect [13] for RuO$_2$ .

In the absence of conclusive evidence as to the precise nature of long-range magnetic order in ruthenium dioxide, we explore three plausible magnetic motifs. All candidates possess an antiferromagnetic order, and differ with regard to magnetic anisotropy. Our results account for rotation of the crystal about the reflection vector, known as an azimuthal angle scan. The chiral signature of charge-magnetic interference in x-ray diffraction is derived from amplitudes for all four polarization channels. Findings suggest that an unambiguous statement concerning the magnetic content of Bragg spot intensities is a fraught task, since charge and magnetic intensities are in quadrature in the rotated channel of polarization. Whereas, the proposed chiral signature is direct evidence of antiferromagnetism. The only relevant x-ray diffraction experiment on a crystal of RuO$_2$ we have found does not exploit circular polarization [8]. Instead, the authors report intensity in the rotated channel of polarization of one basis forbidden Bragg spot as a function of azimuthal angle.

## II. MAGNETIC MOTIFS

The parent crystal structure P4$_2$/mnm (No. 136) is tetragonal and centrosymmetric. Ruthenium ions use sites 2a with symmetry mmm (D$_{2h}$). Cell lengths of RuO$_2$ are a = b ≈ 4.492 Å, c ≈ 3.106 Å [7]. The chemical structure is depicted in Fig. 1.

Three magnetic motifs are considered [14]: (1) tetragonal P4$_2$′/mnm′ (No. 136.499, magnetic crystal class 4′/mm′m, and subgroups include m′m′m and 2/m), (2) orthorhombic Pnn′m′ (No. 58.398, m′m′m, No. 58 is a maximal non-isomorphic subgroup), and (3) monoclinic P2$_1$/c (Nos. 14.75, 2/m). A linear magnetoelectric effect is forbidden in all motifs, while a higher-order HEE-type effect is allowed. Motifs differ with respect to bulk magnetism, which is forbidden in (1) and permitted in (2) and (3). For all motifs, ruthenium ions are in sites 2a with inversion symmetry, which forbids parity-odd (polar) Ru multipoles, and one of the symmetry elements of the space group relates the two sites. Ruthenium multipoles are set in orthogonal Cartesian axes (ξ, η, ζ) that coincide with cell edges for motifs (1) and (2). For the monoclinic structure, we follow convention and equate the ξ-axis with reciprocal lattice vector a*$_m$ ∝ (0, c, −a), together with η ∝ (−a, 0, 0) and ζ ∝ (0, a, c). Vectors a$_m$ ∝ (0, 0, −c) and ζ subtend an angle = 124.66º.

Regarding motifs (1) and (2), the fact that magnetic structure with antiferromagnetic moments within the (ab)-plane of a tetragonal crystal allows ferro-canting while the same magnetic structure with moments along the tetragonal axis does not is quite a common situation. Microscopically, the canting is due to antisymmetric exchange between spins **S**, also known as the Dzyaloshinskii–Moriya interaction, which is given by the mixed product D$_{ij}$ [**S**$_i$ × **S**$_j$] for an exchange bond. Here, the Dzyaloshinskii vector D$_{ij}$ is characteristic of structural distortions of the given crystal. If D$_{ij}$ is along the c-axis then D$_{ij}$ [**S**$_i$ × **S**$_j$] vanishes if the spins

$S_i$ and $S_j$ are also along the c-axis. While if $S_i$ and $S_j$ are in-plane then $D_{ij} [S_i \times S_j]$ is non-zero. The monoclinic magnetic space group $P2_1/c$ that defines motif (3) allows the magnetic moment to be in the (bc)-plane, i.e., a ferro component along the tetragonal a-axis and antiferro components along b- and c-axes.

Parity-even (axial) ruthenium spherical multipoles $\langle T^K_Q \rangle$ have integer rank K and projections Q that obey $-K \leq Q \leq K$ [15] (Cartesian and spherical components of a dipole $R = (x, y, z)$ are related by $x = (R_{-1} - R_{+1})/\sqrt{2}$, $y = i(R_{-1} + R_{+1})/\sqrt{2}$, $z = R_0$). The time signature of a multipole is chosen to be $(-1)^K$, e.g., a dipole is time-odd (magnetic) and a quadrupole is time-even (charge like). The complex conjugate is defined as $\langle T^K_Q \rangle^* = (-1)^Q \langle T^K_{-Q} \rangle$, with a phase convention $\langle T^K_Q \rangle = [\langle T^K_Q \rangle' + i\langle T^K_Q \rangle'']$ for real and imaginary parts labelled by single and double primes, respectively. Angular brackets $\langle ... \rangle$ denote the time-average, or expectation value, of the enclosed tensor operator. For motif (1), site symmetry demands even projections $Q = 2n$ together with $\langle T^K_Q \rangle = (-1)^n \langle T^K_{-Q} \rangle$, while in motif (2) symmetry demands K + Q even and no more. Spatial inversion symmetry alone is required at Ru sites in motif (3). Electronic multipoles can be calculated using standard tools of atomic physics given a suitable wavefunction. Alternatively, multipoles can be estimated from a tried and tested simulation program of electronic structure [16].

The atomic configuration $4d^4$ in a crystal field has been investigated by several authors [7, 17, 18, 19]. In the (slightly tetragonally distorted) octahedral symmetry of the Ru ion the 4d states are split into $t_{2g}$ and $e_g$. Electric dipole transitions from $p_{1/2}$ to $d(t_{2g})$ are symmetry forbidden, but become weakly allowed in the presence of (2p-4d) electrostatic interactions. With the 2p spin-orbit interaction much larger than the other interactions, the core-hole j = 1/2 ($L_2$) and j = 3/2 ($L_3$) levels are not mixed [18]. The $L_2$ absorption peak is mainly due to transitions $2p_{1/2} \rightarrow 4d(e_g)$, with a small shoulder at ~ 1 eV lower photon energy due to transitions into the $4d(t_{2g})$ states. The orbital resolved density of states derived by Berlijn *et al*. [7] is shown in Fig. S3 of their Supplementary Material. The d(xz) and d(yz) are singly occupied, and the $d(x^2 - y^2)$ state (i.e., the d(xy) after 45° rotation about z in the coordinate system) is doubly occupied. Thus, four electrons in $t_{2g}$, in accordance with an octahedral low-spin atomic configuration, because the 4d crystal field interaction is strong (several eV) [19]. The ground-state occupation of the $e_g$ states (blue line in Fig. S3 [7]) is very small.

### III. X-RAY DIFFRACTION

An electronic structure factor for diffraction embodies selection rules imposed by all elements of symmetry in a magnetic space group. We use $\Psi^K_Q = [\exp(i\kappa \cdot d) \langle T^K_Q \rangle_d]$ for a structure factor, where the reflection vector $\kappa$ is defined by integer Miller indices (h, k, l), and the implied sum in $\Psi^K_Q$ is over all Ru sites $d$ in a unit cell of a rutile structure. In an electric dipole - electric dipole absorption event (E1-E1) K = 0, 1, 2. For motifs (1) and (2),

$$\Psi^K_Q(1, 2) = [\langle T^K_Q \rangle + (-1)^{h+k+l} 2_\xi \langle T^K_Q \rangle], \qquad (1)$$

with $2_\xi \langle T^K_Q \rangle = (-1)^K \langle T^K_{-Q} \rangle$. Different site symmetries for Ru ions in (1) and (2) alone distinguishes the motifs. A significantly different result is obtained for motif (3), namely,

$$\Psi^K{}_Q(3) = [\langle T^K{}_Q \rangle + (-1)^{h+k+l} 2_\eta \langle T^K{}_Q \rangle], \qquad (2)$$

with $2_\eta \langle T^K{}_Q \rangle = (-1)^{K+Q} \langle T^K{}_{-Q} \rangle$. Dyad operations $2_\xi$ and $2_\eta$ that occur in Eqs. (1) and (2) are symmetry elements in point groups $D_{4h}$, $D_{2h}$ and $C_{2h}$ that delineate motifs (1), (2) and (3), respectively. Tetragonal Miller indices are displayed in Eq. (2). Basis allowed reflections (K even and Q = 0) are indexed by $h + k + l$ even for all motifs.

The photon scattering length derived from quantum electrodynamics is developed in the small quantity $E/mc^2$, where E is the primary energy ($mc^2 \approx 0.511$ MeV). At the second level of smallness in this quantity the length contains resonant processes that may dominate all other contributions should E match an atomic resonance $\Delta$. Ruthenium $L_2$ and $L_3$ absorption edges occur at energies $\approx 2.97$ keV and $\approx 2.84$ keV, respectively. Assuming that virtual intermediate states are spherically symmetric, to a good approximation, the scattering length $\approx \{F_{\mu\eta}/(E - \Delta + i\Gamma/2)\}$ in the region of the resonance, where $\Gamma$ is the total width of the resonance. The numerator $F_{\mu\eta}$ is an amplitude, or unit-cell structure factor, for Bragg diffraction in the scattering channel with primary (secondary) polarization $\eta$ ($\mu$). By convention, $\sigma$ denotes polarization normal to the plane of scattering, and $\pi$ denotes polarization within the plane of scattering. Fig. 2 depicts polarization states, wavevectors, and the Bragg condition. The illuminated sample is rotated about the reflection vector in an azimuthal angle scan.

Photon and electronic quantities in the scattering amplitude are partitioned in a generalized scalar product $F_{\mu\eta} = \{\mathbf{X}^K \bullet \mathbf{\Psi}^K\}$, with implied sums on rank K [15, 20]. Selection rules on K and Q from the electronic structure factor $\mathbf{\Psi}^K$ are evidently shared with the photon tensor $\mathbf{X}^K$ that defines states of polarization.

Henceforth, we adopt a shorthand ($\mu\eta$) for the scattering amplitude $F_{\mu\eta}$. Scattered intensity picked out by circular polarization in the primary photon beam = $P_2 \Upsilon$ with [9, 10],

$$\Upsilon = \{(\sigma'\pi)^*(\sigma'\sigma) + (\pi'\pi)^*(\pi'\sigma)\}'', \qquad (3)$$

and the Stokes parameter $P_2$ (a purely real pseudoscalar) measures helicity in the primary x-ray beam. Since intensity is a true scalar, $\Upsilon$ and $P_2$ must possess identical discrete symmetries, specifically, both scalars are time-even and parity-odd. Intensity of a Bragg spot in the rotated channel of polarization is proportional to $|(\pi'\sigma)|^2$, and likewise for unrotated channels of polarization.

### IV. SCATTERING AMPLITUDES

Chiral signatures defined by Eq. (3) are created by interference between magnetic dipoles and quadrupoles, which generate Templeton-Templeton scattering [21]. Diffraction enhanced by an E1-E1 absorption event by motifs (1) and (2) at basis forbidden reflections is reported first. Thereafter, discussions of diffraction by motif (3), again with basis forbidden Bragg wavevectors. Results for other Bragg reflections are readily calculated using universal expressions for the four polarization amplitudes in Eq. (3) reported by Scagnoli and Lovesey [20].

### A. Motif (1)

Consider a reflection vector $(2m + 1, 0, 0)$ parallel to the crystal a-axis. The crystal c-axis is chosen to be normal to the plane of scattering for azimuthal angle $\psi = 0$. Scattering amplitudes are $(\sigma'\sigma)_1 = 0$ and,

$$(\pi'\sigma)_1 = -\cos(\theta)\sin(\psi)\,[i\sqrt{2}\,\langle T^1_\zeta\rangle + 2\,\langle T^2_{+2}\rangle''],$$

$$(\sigma'\pi)_1 = -\cos(\theta)\sin(\psi)\,[-i\sqrt{2}\,\langle T^1_\zeta\rangle + 2\,\langle T^2_{+2}\rangle''],$$

$$(\pi'\pi)_1 = i\sqrt{2}\,\sin(2\theta)\cos(\psi)\,\langle T^1_\zeta\rangle,$$

$$\Upsilon(1) = \sqrt{2}\,\sin(2\theta)\cos(\theta)\sin(2\psi)\,\langle T^1_\zeta\rangle\,\langle T^2_{+2}\rangle'', \qquad (2m+1, 0, 0) \qquad (4)$$

where $\theta$ is the Bragg angle depicted in Fig. 2. Amplitudes for $(0, 2m + 1, 0)$ differ from Eq. (4) in the sign of the quadrupole alone. Intensities in rotated channels $|(\pi'\sigma)_1|^2 = |(\sigma'\pi)_1|^2 \propto \sin^2(\psi)$ $[\langle T^1_\zeta\rangle^2 + 2\,(\langle T^2_{+2}\rangle'')^2]$ as a function of azimuthal angle do not distinguish between Templeton-Templeton scattering and magnetic scattering. Next, we consider a reflection vector $(0, 0, l)$ with $l$ odd that is parallel to the dipole moment $\langle T^1_\zeta\rangle$. The crystal a-axis is set normal to the plane of scattering for $\psi = 0$. Amplitudes and the chiral signature are,

$$(\sigma'\sigma)_1 = -2\sin(2\psi)\,\langle T^2_{+2}\rangle'',\quad (\pi'\pi)_1 = \sin^2(\theta)\,(\sigma'\sigma)_1,$$

$$(\pi'\sigma)_1 = \sin(\theta)\,[\,i\sqrt{2}\,\langle T^1_\zeta\rangle - 2\cos(2\psi)\,\langle T^2_{+2}\rangle''],$$

$$(\sigma'\pi)_1 = \sin(\theta)\,[\,i\sqrt{2}\,\langle T^1_\zeta\rangle + 2\cos(2\psi)\,\langle T^2_{+2}\rangle''],$$

$$\Upsilon(1) = \sqrt{2}\,\sin(2\theta)\cos(\theta)\sin(2\psi)\,\langle T^1_\zeta\rangle\,\langle T^2_{+2}\rangle''. \qquad (0, 0, 2m+1) \qquad (5)$$

Note that the formula for $\Upsilon(1)$ is identical to the formula in Eq. (4), while the angle $\theta$ is different for the two Bragg spots. Evidently, the 90° phase shift between the charge and magnetic contributions to $(\pi'\sigma)$ and $(\sigma'\pi)$ allow $\Upsilon(1)$ different from zero. Exactly the same mechanism is present in magnetic motifs (2) and (3).

### B. Motif (2)

Starting with Eq. (1) and appropriate restrictions on K and Q for motif (2), we find $(\sigma'\sigma)_2 = 0$ and,

$$(\pi'\sigma)_2 = -\cos(\theta)\,[i\sqrt{2}\cos(\psi)\,\langle T^1_\eta\rangle + 2\sin(\psi)\,\langle T^2_{+2}\rangle''],$$

$$(\sigma'\pi)_2 = -\cos(\theta)\,[-i\sqrt{2}\cos(\psi)\,\langle T^1_\eta\rangle + 2\sin(\psi)\,\langle T^2_{+2}\rangle''],$$

$$(\pi'\pi)_2 = -i\sqrt{2}\,\sin(2\theta)\sin(\psi)\,\langle T^1_\eta\rangle,$$

$$\Upsilon(2) = -2\sqrt{2}\,\sin(2\theta)\cos(\theta)\sin^2(\psi)\,\langle T^1_\eta\rangle\,\langle T^2_{+2}\rangle''. \qquad (2m+1, 0, 0) \qquad (6)$$

Intensities $|(\pi'\sigma)_1|^2 = |(\sigma'\pi)_1|^2$ can be taken to be functions of $\sin^2(\psi)$ alone. Note that $\Upsilon(1)$ and $\Upsilon(2)$ do not depend on the azimuthal angle $\psi$ in the same way for the same reflection vector. For a reflection vector $(0, 0, 2m + 1)$,

$$(\sigma'\sigma)_2 = -2\sin(2\psi)\langle T^2_{+2}\rangle'',$$

$$(\pi'\sigma)_2 = [i\sqrt{2}\cos(\theta)\cos(\psi)\langle T^1_\eta\rangle - 2\sin(\theta)\cos(2\psi)\langle T^2_{+2}\rangle''], \quad (\sigma'\pi)_2 = -(\pi'\sigma)_2,$$

$$(\pi'\pi)_2 = i\sqrt{2}\sin(2\theta)\sin(\psi)\langle T^1_\eta\rangle + \sin^2(\theta)(\sigma'\sigma)_2,$$

$$\Upsilon(2) = 4\sqrt{2}\cos(\theta)\sin(\psi)[\cos^2(\theta)\sin^2(\psi) - 1]\langle T^1_\eta\rangle\langle T^2_{+2}\rangle''. \quad (0, 0, 2m+1) \quad (7)$$

In Eqs. (6) and (7) the the reflection vector is normal to magnetic dipole $\langle T^1_\eta\rangle$.

### C. Motif (3)

The monoclinic motif (3) is defined by the electronic structure factor Eq. (2). As in foregoing calculations, we consider basis forbidden Bragg spots. Specifically, spots indexed by $(2m + 1, 0, 0)$ and $(0, 2m + 1, 0)$, which are not equivalent. In the first case we find $(\sigma'\sigma)_3 = 0$ and,

$$(\pi'\sigma)''_3 = -\sqrt{2}\cos(\theta)[\sin(\psi)\langle T^1_\zeta\rangle + \cos(\psi)\langle T^1_\xi\rangle], \quad (\sigma'\pi)''_3 = -(\pi'\sigma)''_3,$$

$$(\pi'\sigma)'_3 = 2\cos(\theta)[\cos(\psi)\langle T^2_{+1}\rangle'' + \sin(\psi)\langle T^2_{+2}\rangle'']\}, \quad (\sigma'\pi)'_3 = (\pi'\sigma)'_3,$$

$$(\pi'\pi)_3 = i\sqrt{2}\sin(2\theta)[\cos(\psi)\langle T^1_\zeta\rangle - \sin(\psi)\langle T^1_\xi\rangle],$$

$$\Upsilon(3) = -2\sqrt{2}\sin(2\theta)\cos(\theta)\{[\cos(\psi)\langle T^1_\zeta\rangle - \sin(\psi)\langle T^1_\xi\rangle]$$

$$\times [\cos(\psi)\langle T^2_{+1}\rangle'' + \sin(\psi)\langle T^2_{+2}\rangle'']\}. \quad (2m+1, 0, 0) \quad (8)$$

An azimuthal angle $\psi = 34.66°$ places the c-axis in the plane of scattering. Clearly, an intensity $|(\pi'\sigma)_3|^2$ as a function of $\psi$ does not distinguish between dipole and quadrupole contributions to the scattering amplitude for rotated polarization. Amplitudes for the second reflection vector $(0, 2m + 1, 0)$ are significantly different, namely,

$$(\sigma'\sigma)_3 = 4\cos(\alpha)\sin(\psi)[\cos(\psi)\langle T^2_{+2}\rangle'' + \sin(\alpha)\sin(\psi)\langle T^2_{+1}\rangle''], \quad (0, 2m+1, 0)$$

$$(\pi'\sigma)''_3 = -\sqrt{2}\{\cos(\theta)\sin(\psi)\langle T^1_\xi\rangle + [\sin(\alpha)\cos(\theta)\cos(\psi) + \cos(\alpha)\sin(\theta)]\langle T^1_\zeta\rangle\},$$

$$(\pi'\sigma)'_3 = 2\{[\cos(\alpha)\sin(\theta)\cos(2\psi) + \sin(\alpha)\cos(\theta)\cos(\psi)]\langle T^2_{+2}\rangle''$$

$$+ \sin(\psi)[\sin(2\alpha)\sin(\theta)\cos(\psi) - \cos(2\alpha)\cos(\theta)]\langle T^2_{+1}\rangle''\},$$

$$(\pi'\pi)''_3 = \sqrt{2}\sin(2\theta)\{\cos(\psi)\langle T^1_\xi\rangle - \sin(\alpha)\sin(\psi)\langle T^1_\zeta\rangle\},$$

$$(\pi'\pi)'_3 = 2\{\cos(\alpha)\sin^2(\theta)\sin(2\psi)\langle T^2_{+2}\rangle'' - \sin(2\alpha)[1 - \sin^2(\theta)\sin^2(\psi)]\langle T^2_{+1}\rangle''\}. \quad (9)$$

In this case, the c-axis is in the plane of scattering for $\psi = 0$. The Bragg angle for the reflection vector $(0, 1, 0)$ is $\theta = 27.71°$, and $\alpha = 20.68°$. Fig. 5 includes a simulation of the chiral signature, for which the formula is cumbersome, as a function of azimuthal angle together with intensity $|(\sigma'\sigma)_3|^2$.

### V. CALCULATED INTENSITY AND EXPERIMENTAL DATA

We discuss what can be learned on the basis of our calculations from available experimental data for resonance-enhanced x-ray diffraction from ruthenium dioxide. Figs. 3 and 4 include data reported by Zhu et al. [8] for $|(\pi'\sigma)|^2$ collected at the Bragg spot $(1, 0, 0)$ with signal enhancement coming from the Ru $L_2$ absorption edge. The c-axis is in the plane of

scattering at the origin of their azimuthal angle scan. Turning to our calculated intensities, the azimuthal angle dependence for motifs (1) and (2), namely, $|(\pi'\sigma)_1|^2$ and $|(\pi'\sigma)_2|^2$ in Eqs. (4) and (6), is $\cos^2(\psi)$ and this is not a good representation of the experimental data. Zhu *et al*. [8] are of the same mind. They report a successful quest for a better representation of their data using a model designed to represent a dipole moment that departs from the direction of a crystal axis. Success leads to a claim by the authors to have evidence of long-range magnetic order in ruthenium dioxide. We are not of the same opinion, because charge and magnetic scattering are not in phase as it is assumed to be in their model scattering amplitude. We have been unable to achieve the same high quality of fit between any of our soundly based amplitudes, however. Tetragonal and orthorhombic motifs labelled (1) and (2) fall short in this regard, as already mentioned.

A fit of $|(\pi'\sigma)_3|^2$ derived from Eq. (8) to the data is displayed in Fig. 3. When used for the reflection vector (1, 0, 0), the monoclinic motif (3) predicts intensity as a function of $\psi$ that is identical for dipoles and charge-like quadrupoles, i.e., an azimuthal-angle scan is not sufficient to distinguish between magnetic and Templeton-Templeton scattering in this motif. Out of interest, we made a fit using intensity predicted for the reflection vector (0, 1, 0), Eq. (9), and the result is displayed in Fig. 4. Even with four unknowns, two dipoles and two quadrupoles, the fit is not of good quality.

We propose the chiral signature $\Upsilon$ defined in Eq. (3) as an unambiguous identifier of long-range magnetic order, since it is sourced from the interference of magnetic and Templeton-Templeton scattering. For illustration purposes, Fig. 5 displays a simulation of $\Upsilon(3)$ for the reflection vector (0, 1, 0) using dipole and quadrupoles inferred from the fit in Fig. 4. Included also in Fig. 5 is a simulation of intensity in the unrotated $\sigma'\sigma$ channel of polarization that is difficult to entirely eliminate from intensity collected in the $\pi'\sigma$ channel. Intensity $|(\sigma'\sigma)_3|^2$ calculated from Eq. (9) is zero at the origin of the azimuthal-angle scan,

## VI. DISCUSSION

In summary, we have demonstrated that charge-magnetic interference in resonant x-ray diffraction is most likely a characteristic property of ruthenium dioxide. Experiments of choice use circularly polarized x-rays to measure a chiral signature of the interference, with prior successful outcomes on haematite, chalcopyrite and terbium manganate samples [9, 10]. Charge and magnetic contribution to intensity in the rotated channel of polarization are in quadrature. Our predictions for $RuO_2$ are based on three antiferromagnetic motifs that differ with respect to magnetic anisotropy. Motifs are constructed from a centrosymmetric crystal that presents purely real charge scattering and magnetic order that does not break translation symmetry, together with the absence of a linear magnetoelectric effect (ME) which demands anti-inversion among elements of the magnetic crystal class.

For the moment, the nearest we get to the proposed experimental investigation is a measurement of intensity in the rotated channel of polarization [8]. None of our motifs provide a very good account of the published data, which is reproduced in Figs. (3) and (4). Possible explanations for the shortcoming include leakage from the unrotated channel of polarization, which is impossible for us to quantify. The scattering amplitude used by Zhu *et al*. [8] to successfully analyse their data returns a null chiral signature, because magnetic symmetry in the amplitude is incorrect.

Germane rutile-type transition metal difluorides in the news include $MnF_2$ [22] and $NiF_2$ [23, 24]. These compounds use motifs (1) and (2), respectively. $MnF_2$ is a uniaxial antiferromagnet (easy axis along the c direction), Néel temperature $T_N \approx 66.7$ K, an atomic configuration $Mn^{2+}$ ($3d^5$), and cell lengths a = b ≈ 4.87 Å, c ≈ 3.30 Å. The nickel difluoride displays weak ferromagnetism below $T_N \approx 68.5$ K, with an atomic configuration $Ni^{2+}$ ($3d^8$) and cell lengths a = b ≈ 4.68 Å, c ≈ 3.06 Å. Recently published simulations of electronic structure indicate metastable ferromagnetism in $NiF_2$ below ≈ 10 K [24]. Energies of L-edges of Mn and Ni correspond to photon wavelengths ≈ 20 Å and the Laue condition for diffraction is not satisfied. Measurements of difluoride Bragg diffraction patterns may benefit from intensity enhancement offered by a K-edge resonance, however, that occur at ≈ 6.54 keV and ≈ 8.34 keV for Mn and Ni, respectively. The electric quadrupole - electric quadrupole (E2-E2, 1s → $n$d) process engages orbital magnetism alone in a $n$d ion [25], while the E1-E1 (2p → $n$d) process featured in the main text picks up spin and orbital magnetisms [15, 26]. Results on haematite ($Fe^{3+}$, $3d^5$) reported by Finkelstein *et al*. [27] are thoroughly discussed by Cara and Thole [28], while diffraction patterns gathered at a later date revealed the chiral signature [9]. The time between the publications saw reports of diffraction patterns enhanced by nickel and vanadium K-edges [29-32]. A challenge posed by charge-orbital ordering in mixed valence perovskites was an early beneficiary of a strong resonance at the Mn K-edge [33, 34].

A polarization dependence of neutron scattering is usually described by a departure from unity of the ratio of the reflected intensities for primary neutron beams of opposite polarization [35, 36]. Such a departure in the ratio of intensities is allowed by motifs we consider. For, non-magnetic (nuclear) and magnetic neutron amplitudes possess a like phase and interfere in diffraction. By contrast, there is no neutron polarization dependence in diffraction by chromium sesquioxide ($Cr_2O_3$), because nuclear and magnetic amplitudes are 90º out of phase. Charge and magnetic x-ray amplitudes for this corundum ME material do not differ in phase and our chiral signature is identically zero [15]. Likewise for tetragonal gadolinium tetraboride which is the subject of an Appendix [37].

**ACKNOWLEDGEMENT** Dr Gøran J. Nilsen gave advice on polarized neutron diffraction.

## APPENDIX

Two principal differences between $GdB_4$ and models discussed for $RuO_2$ are the ME and site symmetries. For gadolinium tetraboride presents a linear ME and Gd ions occupy sites 4g in tetragonal P4/m′b′m′ (No.127.395, magnetic crystal class 4/m′m′m′) that are not centres of inversion symmetry, meaning parity-odd Gd multipoles are permitted [37]. Site symmetry demands that a Gd multipole obeys $\langle O^K_Q \rangle = [\exp(i\pi Q/2) (-1)^K \langle O^K_{-Q} \rangle]$, while $[\sigma_\theta \sigma_\pi (-1)^Q]$ = +1 constrains projections Q. Multipoles $\langle T^K_Q \rangle$ encountered in x-ray diffraction enhanced by an E1-E1, or E2-E2, absorption event possess a time signature $\sigma_\theta = (-1)^K$ and parity $\sigma_\pi = +1$. The antiferromagnetic motif of axial magnetic dipoles does not break translation symmetry. Dipoles are confined to the ξ-η plane, and $\langle T^1_\xi \rangle = \langle T^1_\eta \rangle$ while $\langle T^1_\zeta \rangle = 0$. Parity-odd multipoles with ranks K = 1, 2 and 3 observed in diffraction enhanced by an E1-E2 absorption event have the signature $\sigma_\pi = -1$, with $\sigma_\theta \sigma_\pi = +1$ and $\sigma_\theta \sigma_\pi = -1$ for Dirac and polar multipoles, respectively. Notably, symmetry allows the Dirac monopole visible in scattering enhanced by an E1-M1 absorption event [38]. Dirac dipoles (anapoles) are forbidden, whereas a polar dipole $\langle U^1_\xi \rangle = \langle U^1_\eta \rangle$ is allowed [15].

Environments at sites 4g in P4/m′b′m′ are related by dyad rotations about axes ($\xi$, $\eta$, $\zeta$), which coincide with cell edges. In consequence, the electronic structure factor for GdB$_4$ does not depend on $\sigma_\theta$ or $\sigma_\pi$ explicitly. Site symmetry says the sign of $\sigma_\theta \sigma_\pi$ constrains Q, together with the mentioned proportionality factor between $\langle O^K_Q \rangle$ and $\langle O^K_{-Q} \rangle$ created by a dyad symmetry operation parallel to ($\xi + \eta$). We find,

$$\Psi^K_Q(GdB_4) = \langle O^K_Q \rangle \, [(-1)^k \, \Pi_Q(h + k) + \exp(-i\pi Q/2) \, (-1)^h \, \Pi_Q(h - k)],$$

where,

$$\Pi_Q(h \pm k) = \{\exp[i\varphi(h \pm k)] + (-1)^Q \, \exp[-i\varphi(h \pm k)]\}, \tag{A1}$$

and $\varphi = 2\pi x$ with x ≈ 0.317 (a = b ≈ 7.131 Å, c ≈ 4.050 Å) [37]. The chiral signature Eq. (3) derived from Eq. (A1) for basis forbidden reflections is zero for all types of multipoles, because scattering amplitudes are purely real.

By way of an illustration, reflections indexed (2m + 1, 0, 0) have parity-even scattering amplitudes $(\sigma'\sigma)_t = 0$ and,

$(\pi'\sigma)_t = 2 \cos(\theta) \, [\sqrt{2} \cos(\psi) \sin(\varphi \, h) \, \langle T^1_\xi \rangle - 2 \sin(\psi) \cos(\varphi \, h) \, \langle T^2_{+2} \rangle''],$

$(\sigma'\pi)_t = -2 \cos(\theta) \, [\sqrt{2} \cos(\psi) \sin(\varphi \, h) \, \langle T^1_\xi \rangle + 2 \sin(\psi) \cos(\varphi \, h) \, \langle T^2_{+2} \rangle''],$

$(\pi'\pi)_t = 2\sqrt{2} \sin(2\theta) \sin(\psi) \sin(\varphi \, h) \, \langle T^1_\xi \rangle. \tag{A2}$

Evidently, there is a coherent sum of Templeton-Templeton and magnetic dipole contributions to the rotated channel of polarization, with intensity $|(\pi'\sigma)|^2$ a function of $\cos(2\psi)$ and $\sin(2\psi)$. Azimuthal angle scans can reveal information on the quadrupole, with $\xi\eta$-type symmetry, and the dipole, abetted by scans performed on different Bragg spots.

Allowed Dirac multipoles include $\langle G^2_0 \rangle$, $\langle G^2_{+2} \rangle''$ and $\langle G^3_{+2} \rangle'$, but $\langle G^2_0 \rangle$ does not contribute to the diffraction amplitude for h or k odd. With a reflection vector (2m + 1, 0, 0) the amplitude in the rotated channel of polarization for an E1-E2 event is [20],

$(\pi'\sigma)_g = (2/\sqrt{15}) \sin(2\theta) \cos(\psi) \cos(\varphi \, h) \, [2\sqrt{2} \, \langle G^2_{+2} \rangle'' - (3 \cos(2\psi) - 1) \, \langle G^3_{+2} \rangle']. \tag{A3}$

While non-zero $\langle G^K_Q \rangle$ depend on magnetic order, polar multipoles $\langle U^K_Q \rangle$ arise from absence of inversion symmetry in sites occupied by Gd ions. A factor $[\sin(2\theta) \sin(\psi) \sin(\varphi \, h)]$ is common to all multipoles in the purely real amplitude $(\pi'\sigma)_u$ [20]. Participating polar multipoles are $\langle U^1_\xi \rangle = -\sqrt{2} \, \langle U^1_{+1} \rangle'$, $\langle U^2_{+1} \rangle'$, $\langle U^3_{+1} \rangle'$ and $\langle U^3_{+3} \rangle'$.

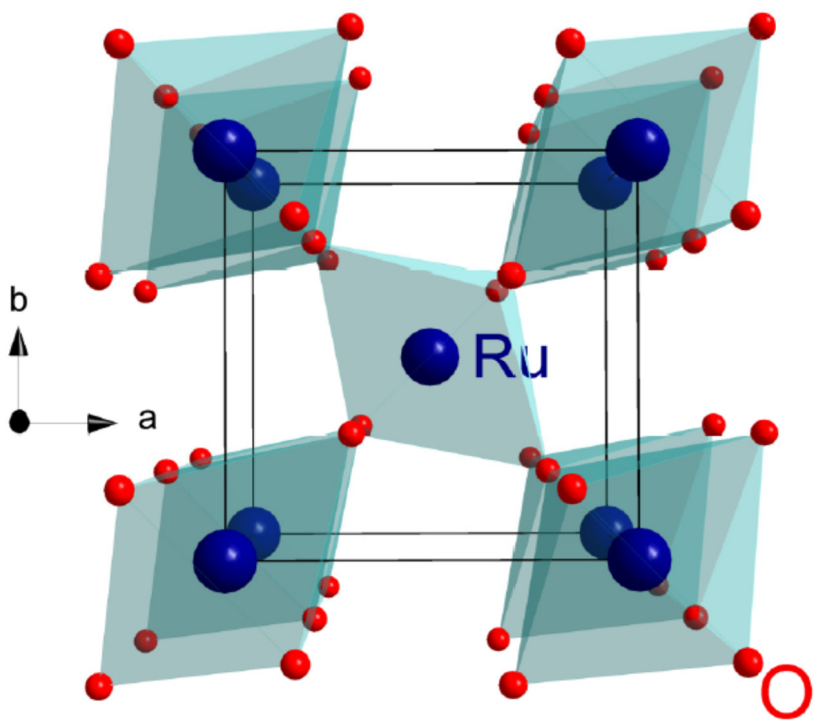

Fig. 1. Crystal structure of ruthenium dioxide with Ru ions using sites 2a in P4$_2$/mnm (No. 136).

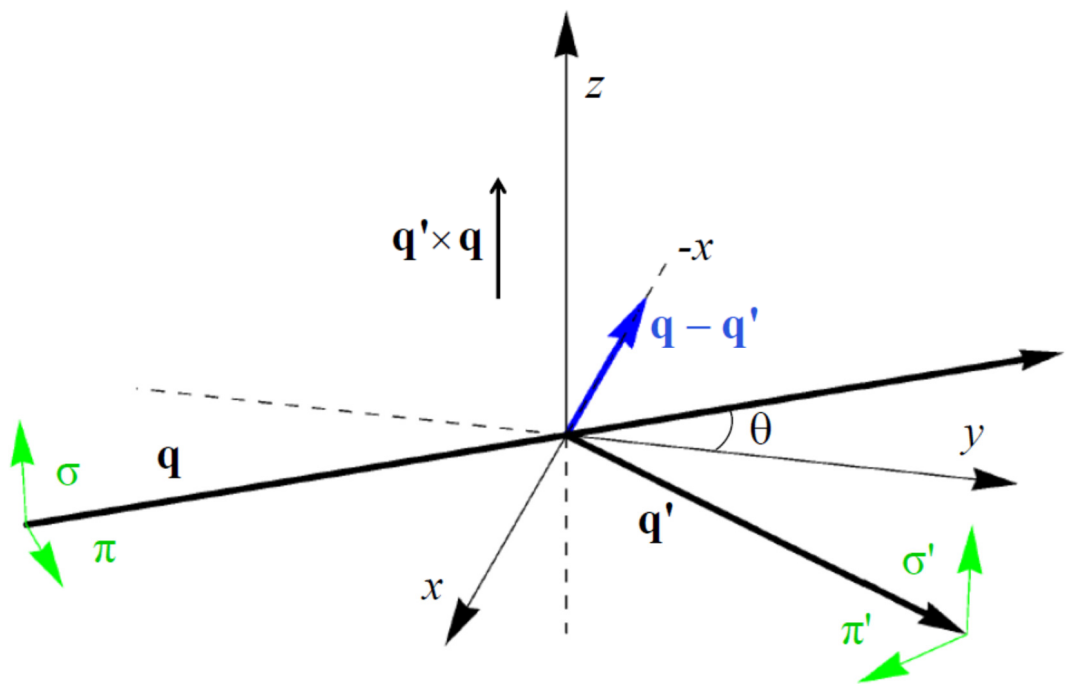

FIG. 2. X-ray diffraction. Primary (σ, π) and secondary (σ′, π′) states of polarization.

Corresponding wavevectors **q** and **q'** subtend an angle 2θ, and the reflection vector **κ** = **q** − **q'**. Orthonormal coordinates (ξ, η, ζ) for the tetragonal (1), orthorhombic (2) or monoclinic (3) magnetic motifs and depicted co-ordinates (x, y, z) coincide in the nominal setting of the crystal.

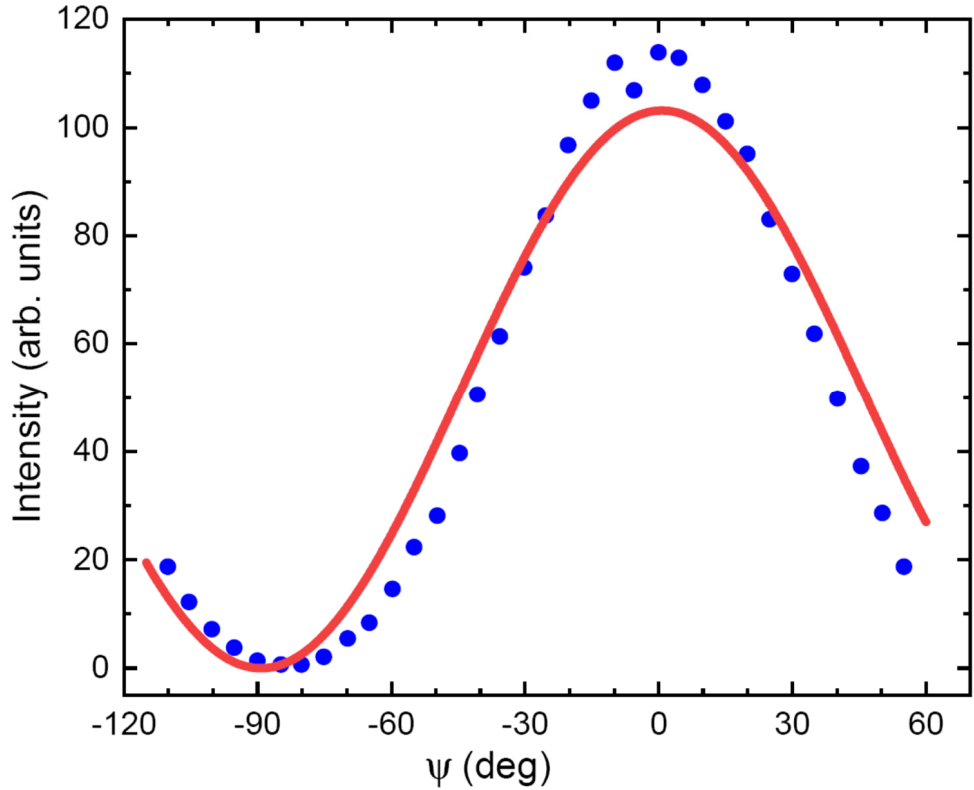

FIG. 3. Red curve: intensity $|(\pi'\sigma)_3|^2$ using a reflection vector (1, 0, 0) and amplitude Eq. (8). Blue dots: experimental data from Ref. [8]. Curve is generated with $[8.28 \cos(\psi + 34.66) + 5.88 \sin(\psi + 34.66)]^2$ and inferred values of the two dipoles and two quadrupoles are $\langle T^1_\zeta \rangle$ or $\langle T^2_{+2} \rangle'' \approx 5.88$ and $\langle T^1_\xi \rangle$ or $\langle T^2_{+1} \rangle'' \approx 8.28$. Multipoles in the monoclinic motif (3) are set in orthonormal coordinates (ξ, η, ζ) defined with respect to the tetragonal host using ξ ∝ (0, c, −a), η ∝ (−a, 0, 0) and ζ ∝ (0, a, c).

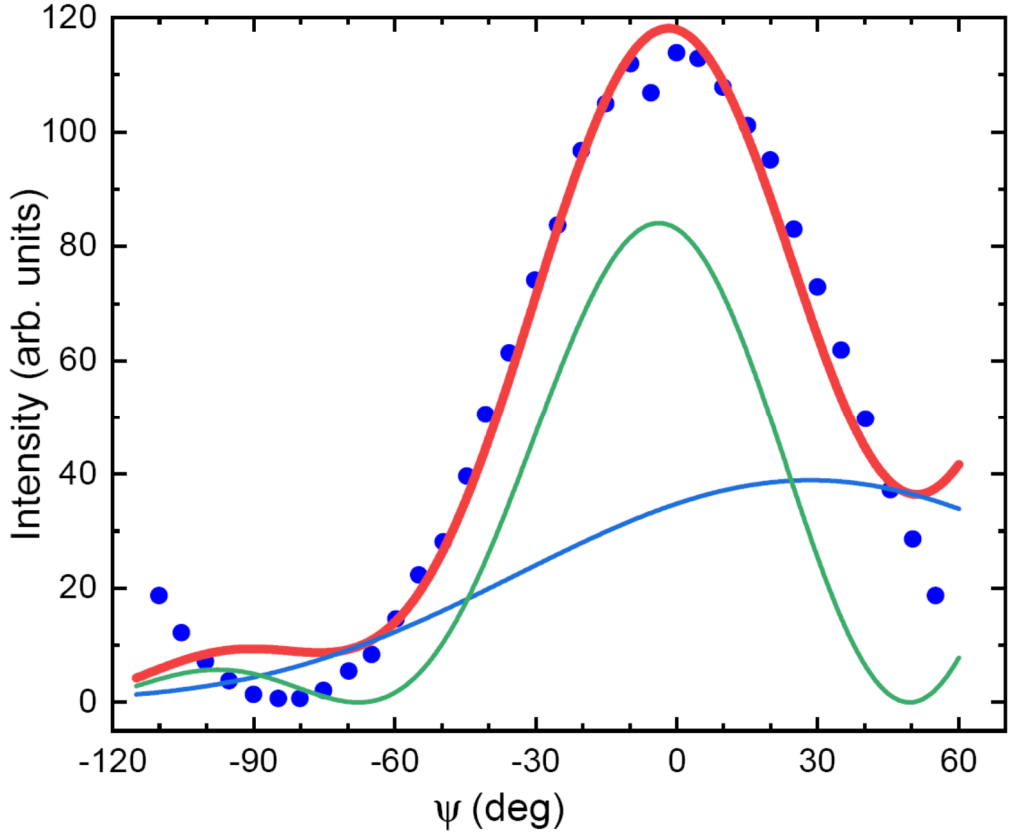

FIG. 4. Red curve: intensity $|(\pi'\sigma)_3|^2$ with the amplitude Eq. (9) for motif (3). Green curve: quadrupole (Templeton-Templeton) contribution. Blue curve: magnetic dipole contribution. Experimental data from Ref. [8], as in Fig. 3. Inferred values of multipoles are $\langle T^1\xi \rangle \approx 1.07$, $\langle T^1\zeta \rangle \approx 5.58$, $\langle T^2_{+2} \rangle'' \approx 6.10$ and $\langle T^2_{+1} \rangle'' \approx 2.31$. Dipoles $\langle T^1\xi \rangle$ and $\langle T^1\zeta \rangle$ are aligned with (0, c, −a) and (0, a, c) in tetragonal $P4_2/mnm$ (No. 136), respectively.

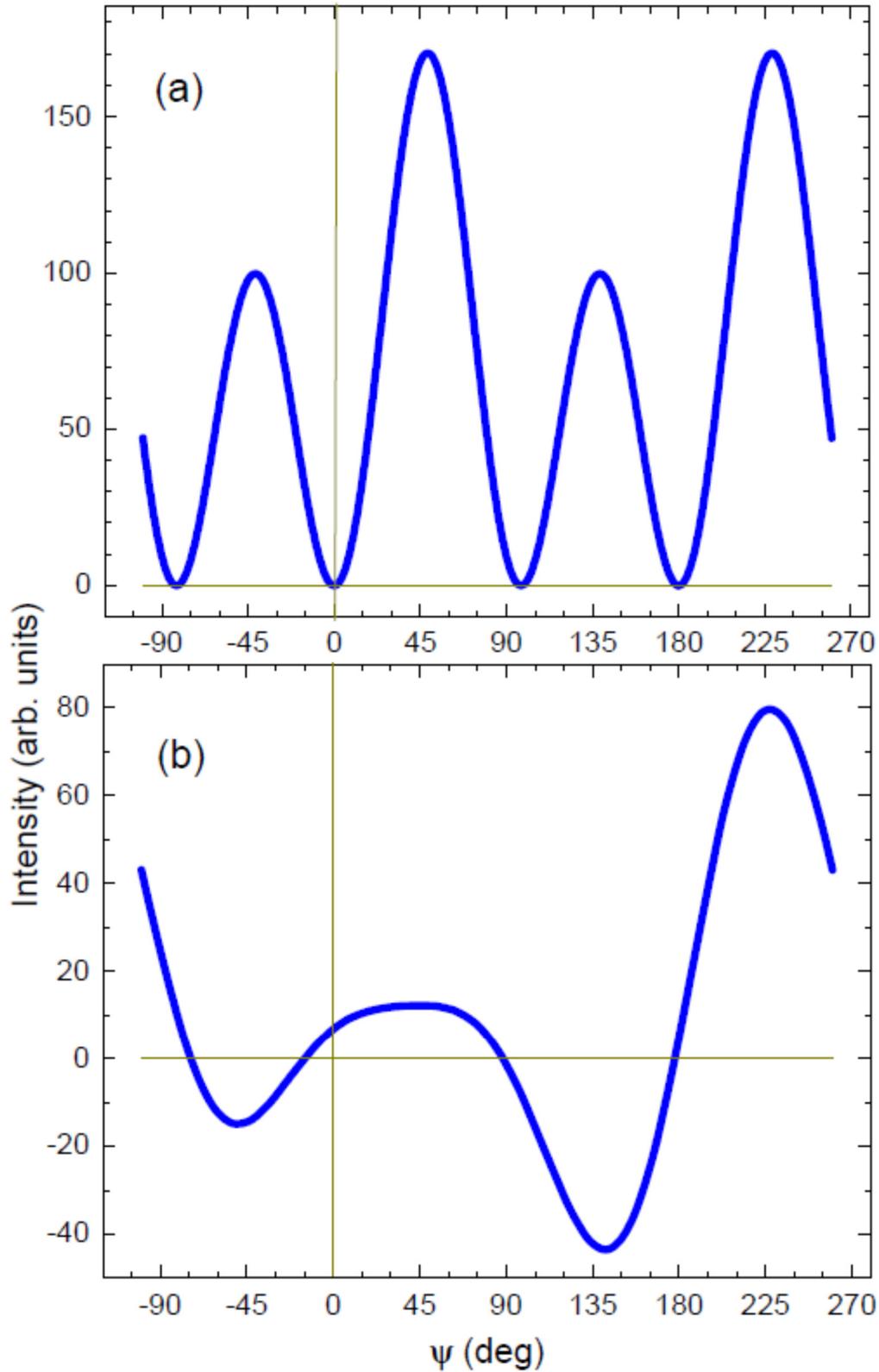

FIG. 5. Panel (a) simulations of intensity $|(\sigma'\sigma)_3|^2$ and panel (b) chiral signature $\Upsilon(3)$ for the Bragg spot (0, 1, 0) as a function of azimuthal angle $\psi$ in the range $-100°$ to $+260°$. Values assigned to multipoles in Eq. (9) are listed in the caption to Fig. 4, and they are inferred from a fit to data reported in Ref. [8].